\documentclass[aps,prd]{revtex4}  % for review and submission
\usepackage[utf8]{inputenc}
\usepackage{graphicx}  % needed for figures
\usepackage{dcolumn}   % needed for some tables
\usepackage{bm}        % for math
\usepackage{amssymb}   % for math
\usepackage{amsmath} % AMS Math Package

\newcommand{\di}{\textrm{d}} % for differentials
\newcommand{\ee}{\textrm{e}} % for e
\newcommand{\ii}{\textrm{i}} % for i
\newcommand{\n}{\phantom{\alpha}} % phantom indices

\newcommand{\marrow}[5]{
    \fmfcmd{style_def marrow#1
    expr p = drawarrow subpath (1/4, 3/4) of p shifted 6 #2 withpen pencircle scaled 0.4;
    label.#3(btex #4 etex, point 0.5 of p shifted 6 #2);
    enddef;}
    \fmf{marrow#1,tension=0}{#5}}       % momentum lines in feynmf

\usepackage{slashed}  % Feynman slash notation (\slashed{D})
\usepackage{feynmp-auto}    % Feynman diagrams
\usepackage{subfigure}

\begin{document}

% The following information is for internal review, please remove them for submission
%\widetext
%\leftline{Version xx as of \today}
%\leftline{Primary authors: Joe E. Physics}
%\leftline{To be submitted to (PRL, PRD-RC, PRD, PLB; choose one.)}
%\leftline{Comment to {\tt d0-run2eb-nnn@fnal.gov} by xxx, yyy}
%\centerline{\em D\O\ INTERNAL DOCUMENT -- NOT FOR PUBLIC DISTRIBUTION}

% the following line is for submission, including submission to the arXiv!!
%\hspace{5.2in} \mbox{Fermilab-Pub-04/xxx-E}

\title{Constraints on hidden gravitons from fifth-force experiments and stellar energy loss}
%\input author_list.tex       % D0 authors (remove the first 3 lines
                             % of this file prior to submission, they
                             % contain a time stamp for the authorlist)
                             % (includes institutions and visitors)
\author{J.\,A.\,R.\,Cembranos\footnote{cembra@ucm.es}, 
        A.\,L.\,Maroto\footnote{maroto@ucm.es}
        and H.\,Villarrubia-Rojo\footnote{hectorvi@ucm.es}}
\affiliation{Departamento de  F\'{\i}sica Te\'orica I, Universidad Complutense de Madrid, E-28040 Madrid, Spain}

\date{\today}

\begin{abstract}
We study different phenomenological signatures associated with new spin-2 particles.
These new degrees of freedom, that we call hidden gravitons, arise in different 
high-energy theories such as extra-dimensional models or extensions of General Relativity.
At low energies, hidden gravitons can be generally described by the Fierz-Pauli Lagrangian.
Their phenomenology is parameterized by two dimensionful constants: their mass and their
coupling strength. In this work, we analyze two different sets of constraints. On the one 
hand, we study potential deviations from the inverse-square law on solar-system 
and laboratory scales. To extend the constraints to scales where the laboratory probes are not 
competitive, we also study consequences on astrophysical objects.
We analyze in detail the processes that may take place in stellar interiors and lead to 
emission of hidden gravitons, acting like an additional source of energy loss. 
\end{abstract}

\maketitle

\section{\label{sec:intro}Introduction}
    Gravity and electromagnetism are, as far as we know today, the only macroscopic forces in nature. Their long-range character can be explained
    according to the massless character of the graviton and the photon. This property, in its turn, is usually justified as a result of the local symmetries
    of both theories, diffeomorphism and $U(1)$ gauge invariance.       
    Nonetheless, it is natural to ask whether they are exactly massless or they just have small masses, and, as a matter of fact, there have been a lot of 
    efforts over the years to test this assumption. On the experimental side, several bounds have been established for a non-zero mass \cite{Goldhaber:2008xy} 
    while on the theoretical side great efforts have been invested in constructing consistent models of massive gravity and massive electrodynamics.
    The starting point of massive electrodynamics is the Proca Lagrangian. It consists on the usual Maxwell Lagrangian plus a simple mass term, that
    explicitly violates the gauge invariance of the theory. This effective approach can be completed at high energies
    through the Stuckelberg or the Higgs mechanisms. On the phenomenological side, one important application of massive electrodynamics has been
    the proposal of a new hypothetical field, known as hidden photon. This hidden photon has associated a large amount of potential
    experimental signatures. In particular, it constitutes a viable candidate for dark matter, whose effects have been explored extensively in the literature 
    \cite{Nelson:2011sf, Arias:2012az, Goodsell:2009xc, Bullimore:2010aj, Abel:2008ai, Cembranos:2016ugq, Cembranos:2012kk}.
    
    On the other hand, massive gravity is usually introduced by using the Fierz-Pauli action \cite{Fierz:1939ix}. It consists on the linearized action from general 
    relativity (GR) plus a suitably chosen mass term. It is worth noting that both the kinetic and the mass term of this action can be derived without previous 
    knowledge of GR. They can be constructed as the most general choices for spin-2 particles, just requiring the absence of ghosts \cite{deRham:2014zqa}.
    This Lagrangian has been thoroughly studied and, today, its properties are well-known and understood. For example, although the free
    action is consistent, a paradoxical behaviour appears when we turn on the interaction. It was discovered independently in
    \cite{Iwasaki:1971uz,vanDam:1970vg,Zakharov:1970cc} that this theory is not continuous in the massless limit, this is the so-called vDVZ discontinuity: 
    the $m=0$ and $m\to 0$ theories are not physically equivalent. 
    The problem of the mass discontinuity can be traced back to the number of degrees 
    of freedom that both theories propagate. While a massless spin-2 particle has only two degrees of freedom (two tensor modes), a massive spin-2 particle has 5 
    (two tensor modes, two vectors and one scalar). It can be shown \cite{deRham:2014zqa,Hinterbichler:2011tt} that when we take the $m\to 0$ limit, the scalar mode 
    becomes strongly coupled, invalidating the linear theory. In fact, when non-linear effects are taken into account, the zero-mass discontinuity is cured through 
    the so-called Vainshtein mechanism \cite{Vainshtein:1972sx}. When the problem of the vDVZ discontinuity seemed solved, Boulware and Deser \cite{Boulware:1973my} 
    showed that for a broad range of extensions of the theory, these non-linear effects also
    introduce a sixth degree of freedom, that turns out to be a ghost (BD ghost). Constructing a fully non-linear, consistent, theory of massive 
    gravity is a big challenge and only very recently it has been possible to evade the BD ghost. In 2010 de Rham, Gabadadze and Tolley (dRGT) constructed 
    a ghost-free non-linear completion of the Fierz-Pauli action, known as ghost-free or dRGT massive gravity \cite{deRham:2010kj}. The dRGT action 
    contains parameters fixing the self-interactions and a reference metric. Shortly after, Hassan and Rosen \cite{Hassan:2011vm} reformulated the
    theory and made this reference metric dynamical. This new formulation is a bimetric theory of gravity, describing at the linear level the 
    evolution of a standard massless graviton plus a massive one, with a Fierz-Pauli mass term. This linearized version of bimetric gravity coincides 
    with the model we will analyze in this work, i.e. massless gravity plus a single massive graviton. For a specialized review on bimetric theory see
    \cite{Schmidt-May:2015vnx}.
    Another context where massive gravitons naturally appear is in extra-dimensional theories of gravity, like the ADD model \cite{ArkaniHamed:1998rs}. 
    In this model, the standard model fields are confined to a 4-brane, while the gravitons (described by the usual Einstein-Hilbert action) can 
    explore a number $n$ of extra large dimensions. When duly compactified, the existence of this new dimensions leads to a tower of Kaluza-Klein (KK) 
    excitations of the graviton. The weak interaction of this KK modes can be compensated by their huge multiplicity and lead to 
    significant deviations from usual gravity. A number of ways to test the model were suggested by the original authors \citep{ArkaniHamed:1998nn} 
    and the experimental constraints were derived in detail in many references \cite{Barger:1999jf,Cullen:1999hc,Hanhart:2000er,Hall:1999mk,
    Hannestad:2001jv,Hannestad:2001xi,Hannestad:2003yd}.
    
    We shall not assume any particular framework for our study. In our model, we will add a single massive graviton to the known particles 
    (linearized bimetric gravity), explore its phenomenological consequences and use the observational evidence to constrain its mass and coupling
    to other fields. In fact, we will employ methods that have become standard to test the impact of
    new light, weakly interacting particles: fifth-force tests and astrophysical energy-loss arguments. They 
    have been applied not only to KK gravitons,
    but also to hidden photons \cite{Dent:2012mx, An:2013yfc}, sterile neutrinos \cite{Raffelt:1990yz, Arguelles:2016uwb} and specially to axions 
    \cite{Krauss:1984gm, Raffelt:1985nk, Pantziris:1986dc, Raffelt:1989zt}.
    
    This paper is organized as follows: in section \ref{sec:massive} we will present the model, the simple Fierz-Pauli Lagrangian, and all the 
    relevant results for the subsequent calculations. Section \ref{sec:fifth} explores the simplest observational consequence of the model, the 
    existence of a fifth force, and use the available experimental data to constrain the mass and coupling of the hidden gravitons. Section 
    \ref{sec:astro} is devoted to astrophysical consequences. It covers some of the processes that may take place inside the stars and induce
    a thermal emission of hidden gravitons. Using astrophysical arguments we can set limits to the efficiency of this novel form of energy loss. These
    limits will allow us to set bounds on the mass and coupling of the hidden gravitons, complementary to those of fifth-force probes. Finally, 
    section \ref{sec:conclusions} collects the main conclusions of the analysis, presents the final exclusion curves and discusses prospects for future work.
    
\section{\label{sec:massive}Massive gravity. Formalism}
    We will start with the Fierz-Pauli Lagrangian, with {\it mostly plus} metric signature ($-$,$+$,$+$,$+$),
    \begin{equation}
        \mathcal{L} =-\frac{1}{2}\partial^{\alpha}h^{\mu\nu}(\partial_\alpha h_{\mu\nu}-2 \partial_{(\mu}h_{\nu )\alpha}
                - \partial_\alpha h \eta_{\mu\nu}+2 \partial_{(\mu}h\eta_{\nu )\alpha})-\frac{1}{2}m^{2}(h_{\mu\nu}^{2}-h^{2})\ ,
    \end{equation}
    that describes a spin-2 particle with mass $m$ on a Minkowski geometry. The kinetic term here is the same appearing in the linearized Einstein-Hilbert action from GR, but
    in fact no previous knowledge of GR is needed to build this Lagrangian. As shown for example in \cite{deRham:2014zqa}, both the kinetic and the 
    mass term are fixed just by requiring the absence of ghosts.
    
    Starting with this Lagrangian, we will construct our free field theory. First, let us rewrite it as 
    \begin{align}
        \mathcal{L} &=\frac{1}{2}h^{\mu\nu}\mathcal{O}^{\alpha\beta}_{\n\n\mu\nu}h_{\alpha\beta}\ ,\\
        S &=\int\di^{4}x\ \mathcal{L}\ ,    
    \end{align}
    where we have integrated by parts and defined the operator
    \begin{equation}
        \mathcal{O}^{\alpha\beta}_{\n\n\mu\nu}=(\delta^{\alpha}_{(\mu}\delta^{\beta}_{\nu )}-\eta_{\mu\nu}\eta^{\alpha\beta})(\square -m^{2})
                -2\delta^{(\alpha}_{(\mu}\partial_{\nu )}\partial^{\beta )}+\eta^{\alpha\beta}\partial_\mu\partial_\nu+\eta_{\mu\nu}\partial^{\alpha}
                \partial^{\beta}\ .
    \end{equation}
    The equations of motion are
    \begin{equation}\label{massive:eom1}
        \frac{\delta S}{\delta h_{\mu\nu}}=0\quad \to\quad \mathcal{O}^{\mu\nu}_{\n\n\alpha\beta}h^{\alpha\beta}=0\ ,
    \end{equation}
    which after a few manipulations can be cast in the form
    \begin{align}
        (\Box-m^{2})h_{\mu\nu} &= 0\ , \label{massive:eom2}\\
        \partial^{\mu}h_{\mu\nu} &=0\ , \label{massive:eom3}\\
        h=h^{\mu}_{\n\mu} &= 0\ . \label{massive:eom4}
    \end{align}
    This is the usual Klein-Gordon equation for a symmetric (10 degrees of freedom, dof), transverse ($-4$ dof), traceless ($-1$ dof) tensor field, 
    describing a total of 5 propagating dof. This naive count of degrees of freedom is supported by a full Hamiltonian analysis 
    \cite{Hinterbichler:2011tt}. Contrary to what happens in linearized GR, which owing to the linearized diffeomorphism invariance only propagates 
    2 tensor modes, in massive gravity we have two tensor modes, two vector modes and one scalar mode.
    
    The solution of \eqref{massive:eom2} can be written as 
    \begin{equation}
        h^{\mu\nu}(x)=\int \frac{\di^{3}\bm{p}}{(2\pi)^{3}2E_{\bm{p}}}\sum_\lambda \Big[a_{\bm{p},\lambda}\epsilon^{\mu\nu}(\bm{p},\lambda)\ee^{\ii px}+
            a^{\dagger}_{\bm{p},\lambda}\epsilon^{\mu\nu *}(\bm{p},\lambda)\ee^{-\ii px}\Big]\ ,
    \end{equation}
    where the polarization tensors satisfy
    \begin{align}
        &p_\mu \epsilon^{\mu\nu}(\bm{p},\lambda)=0\ ,\\
        &\eta_{\mu\nu}\epsilon^{\mu\nu}(\bm{p},\lambda)=0\ ,\\
        &\epsilon^{\mu\nu}(\bm{p},\lambda)\epsilon^{*}_{\mu\nu}(\bm{p},\lambda ')=\delta_{\lambda\lambda '}\ ,\\
        &\sum_\lambda \epsilon^{\mu\nu}(\bm{p},\lambda)\epsilon^{\alpha \beta *}(\bm{p},\lambda)=\frac{1}{2}
            (P^{\mu\alpha}P^{\nu\beta}+P^{\mu\beta}P^{\nu\alpha})-\frac{1}{3}P^{\mu\nu}P^{\alpha\beta}\ , \label{polarization_sum}
    \end{align}
    with $P^{\alpha\beta}=\eta^{\alpha\beta}+p^{\alpha}p^{\beta}/m^{2}$. Since we will only work with conserved sources $\partial T=0$, our scattering
    amplitudes will have the property $p^\mu \mathcal{A}_{\mu\cdots}=0$, so we can make the identification $P_{\mu\nu}\to\eta_{\mu\nu}$ in 
    \eqref{polarization_sum} and work with the sum over polarizations given by
    \begin{equation}
        S^{\mu\nu\alpha\beta}=\frac{1}{2}
            (\eta^{\mu\alpha}\eta^{\nu\beta}+\eta^{\mu\beta}\eta^{\nu\alpha})-\frac{1}{3}\eta^{\mu\nu}\eta^{\alpha\beta}\ , \label{polarization_sum2}
    \end{equation}
    which differs from GR in the factor $1/3$,  owing to the scalar mode contribution. Next, to find the propagator we need to solve
    \begin{equation}
        \mathcal{O}^{\alpha\beta |\sigma\lambda}(p)D_{\sigma\lambda |\mu\nu}(p)=\ii \delta^{\alpha}_{(\mu}\delta^{\beta}_{\nu )}\ .
    \end{equation}
    The solution, as can be checked by direct substitution, is
    \begin{equation}\label{propagator}
        D_{\alpha\beta |\mu\nu}= \frac{-\ii}{p^{2}+m^{2}}\Big[P_{\alpha (\mu}P_{\nu )\beta}-\frac{1}{3}P_{\alpha\beta}P_{\mu\nu}\Big]\ ,
                \qquad\quad P_{\mu\nu}\equiv \eta_{\mu\nu}+\frac{p_\mu p_\nu}{m^{2}}\ .
    \end{equation}
    Now, we need to turn on the interaction. Although we will be more precise about the form of the interaction in section \ref{sec:astro}, for now let 
    us choose a generic source $T_{\mu\nu}$. The Lagrangian with a linear interaction with the source is
    \begin{equation}
        \mathcal{L}=\frac{1}{2}h^{\mu\nu}\mathcal{O}^{\alpha\beta}_{\n\n\mu\nu}h_{\alpha\beta} +\kappa h_{\mu\nu}T^{\mu\nu}\ ,
    \end{equation}
    and the equations of motion are
    \begin{equation}
        \mathcal{O}^{\mu\nu}_{\n\n\alpha\beta}h^{\alpha\beta}=-\kappa T^{\mu\nu}\ ,
    \end{equation}
    with solution
    \begin{equation}
        h_{\mu\nu}(x)=\ii\kappa \int \di^{4}x'\ D_{\mu\nu |\alpha\beta}(x-x')T^{\alpha\beta}(x')\ .
    \end{equation}
    We have presented here the formal developments and results for a linear theory of massive gravity. In the next sections we will explore the
    observational impact of a new massive spin-2 particle, in addition to the usual massless graviton.
    
\section{\label{sec:fifth}Fifth-force constraints}
    \subsection{Theory}
        The first phenomenological conclusion we can extract from the model above is the existence of a new force. In order to see the effect of this new 
        force between two matter particles, e.g. two electrons, one could first compute the one graviton exchange amplitude, then take the 
        non-relativistic limit and identify the interaction potential via the Born approximation. A textbook example can be found in 
        \cite{Peskin:1995ev}. This is the standard procedure when particles with non-trivial parity, like pseudoscalars, are present and mediate spin-dependent 
        forces. See \cite{Moody:1984ba} for an analysis of the axion case and \cite{Fischbach:1999bc} for a discussion of spin-dependent forces.
        However, in our case, to reproduce the results at lowest order it is easier to compute the classical interaction potential.
        
        In the next section we will discuss how this hidden graviton couples to other fields. For now, to compute the macroscopic force that it may 
        produce, we will consider the force mediated between two classical, non-relativistic sources with energy-momentum tensor
        \begin{equation}
            T^{\mu\nu}_{(i)}=M_i \delta^{\mu}_0 \delta^{\nu}_0\delta^{3}(\bm{x}-\bm{x}_i),\qquad i=1,2
        \end{equation}
        i.e. two lumps of matter sitting at $\bm{x}_1$ and $\bm{x}_2$. The interaction potential is
        \begin{align*}
            V   &=-\kappa \int\di^{3}\bm{x}\ h_{\mu\nu}(x)T^{\mu\nu}_2(x)= -\ii \kappa^{2}\int\di^{3}\bm{x}\int\di^{4}y\ T^{\alpha\beta}_1(y)
                    D_{\alpha\beta|\mu\nu}(x-y)T^{\mu\nu}_2(x)\\
                &=-\ii\kappa^{2}M_1M_2 \int\di y^{0}\int\frac{\di^{4}p}{(2\pi)^{4}}D_{0000}(p)\ee^{\ii p^{0}(x_0-y_0)}\ee^{-\ii\bm{p}
                    (\bm{x}_1-\bm{x}_2)}\\
                &=-\ii\kappa^{2}M_1M_2\int\frac{\di^{3}\bm{p}}{(2\pi)^{3}}D_{0000}(p^{0}=0,\bm{p})\ee^{-\ii\bm{p}(\bm{x}_1-\bm{x}_2)}\ .
        \end{align*}
        Now, it is worth recalling the form \eqref{propagator} of the propagator. For massless gravity one can also derive the propagator, after properly 
        fixing the gauge, and the result is the same as in the massive case, save for a factor $1/2$ instead of $1/3$ \cite{Hinterbichler:2011tt}. For the 
        moment, we write the generic form
        \begin{equation}
            \ii D_{0000}(p^{0}=0,\bm{p})=\frac{1-\alpha}{\bm{p}^{2}+m^{2}}\ ,
        \end{equation}
        where $\alpha=1/2,1/3$ for massless/massive gravitons. After performing the integral, we obtain what is to be expected from a massive, even 
        spin, boson: a universally attractive Yukawa force
        \begin{equation}
            V=-\kappa^{2}M_1M_2 \frac{\ee^{-mr}}{4\pi r}(1-\alpha),\qquad r=|\bm{x}_1-\bm{x}_2|\ .
        \end{equation}
        The standard Newtonian potential is recovered in the massless case ($m=0$, $\alpha=1/2$, $\kappa =1/M_\text{Pl}=\sqrt{8\pi G}$) while in the
        massive case we have ($\kappa=1/M_h=\sqrt{8\pi G_h}$)
        \begin{equation}
            V=-\frac{4}{3}G_h M_1 M_2 \frac{\ee^{-mr}}{r}\ .
        \end{equation}
        The appearance of the factor $4/3$ may seem surprising. In fact, it could be reabsorbed in the definition of $G_h$, so that the $m=0$ and
        $m\to 0$ cases will give the same physical results with the identification $G=\frac{4}{3}G_h$. However, this kind of factors reappear when 
        calculating the deflection of light \cite{deRham:2014zqa}. In that case, the factors cannot be reabsorbed, yielding unambiguosly different
        results. As commented in the introduction, this is the vDVZ discontinuity in the massless limit.
        
        So we will stick to this definition of the coupling constant, without reabsorbing the factor $4/3$. The total potential produced by standard
        gravity and this hypothetical new mediator is
        \begin{equation}\label{eq:fifth:final_potential}
            V(r)=-\frac{G M_1M_2}{r}\Big(1+\frac{4}{3}\frac{G_h}{G}\ee^{-mr}\Big)\ .
        \end{equation}
        With this result, we are ready to constrain the possible values of $G_h$ and $m$ using the available data.
        
    \subsection{Experiments}
        Over the last decades there has been an ongoing effort to measure possible deviations from the inverse square law (ISL), without success
        so far. As a result of this effort, there exists a good deal of experimental data, ranging from microscopic to solar-system scales, that
        can be used to put stringent bounds to our model.
        
\begin{figure}[!htb]
                \subfigure{
                    \includegraphics[scale=0.52]{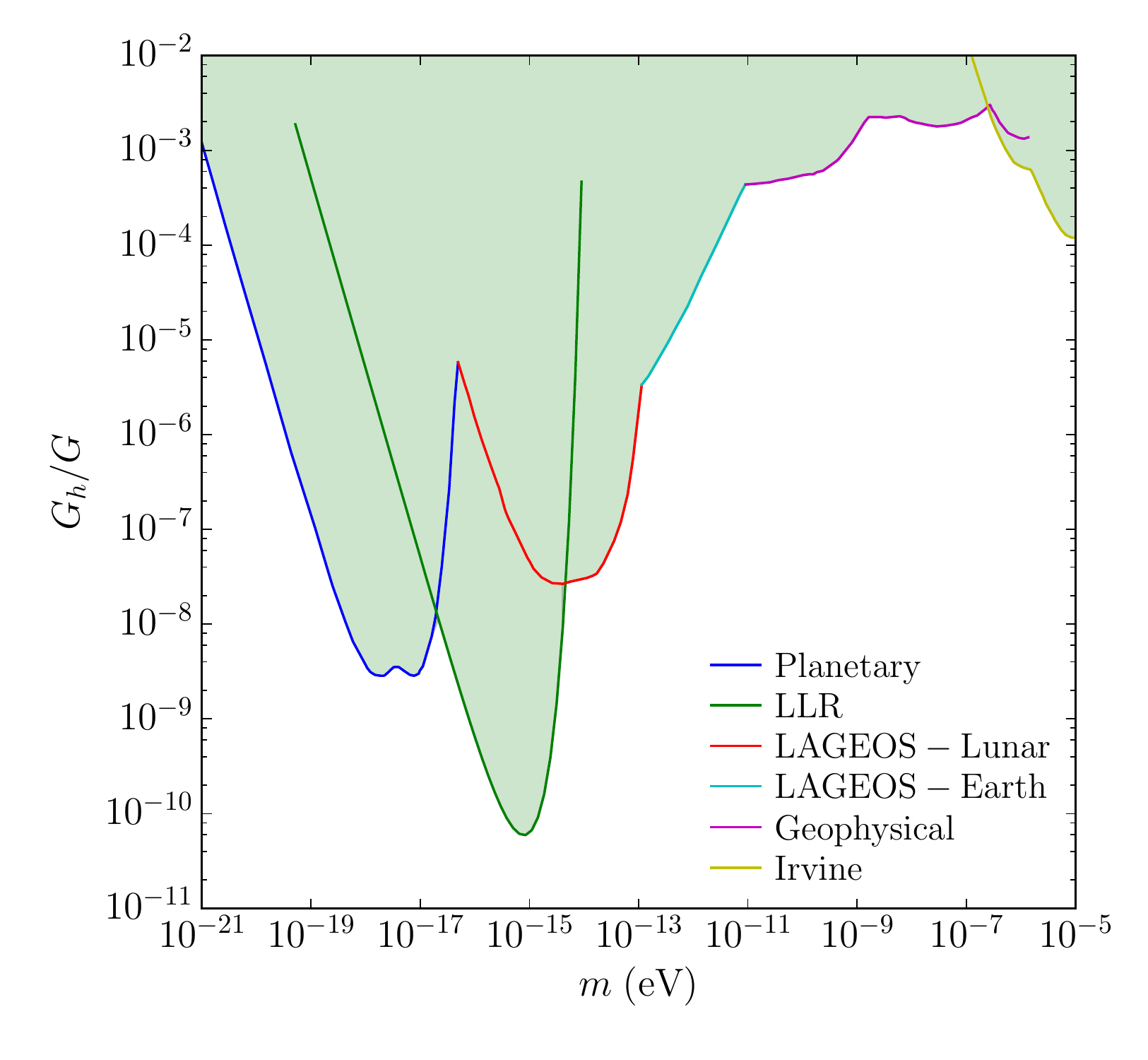}
                }
                \subfigure{
                    \includegraphics[scale=0.52]{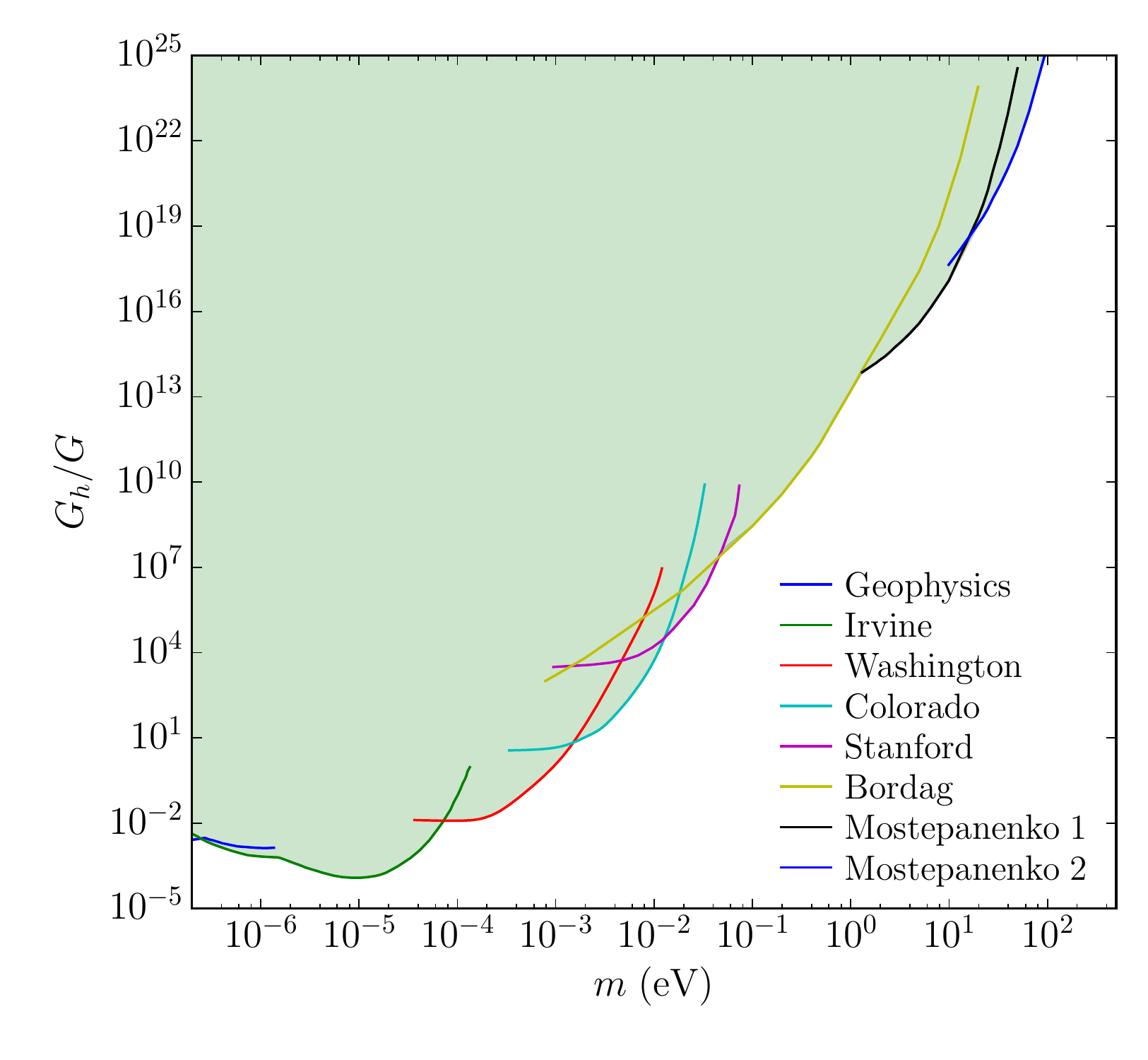}
                }
                \caption{Constraints on the hidden-graviton mass and coupling $G_h$, relative to the standard-graviton coupling. The shadowed 
                    region is excluded by fifth-force tests. The curves have been adapted from the following references: planetary, LLR, LAGEOS-Lunar,
                    LAGEOS-Earth, geophysics \citep{Adelberger:2003zx}, Irvine \cite{Hoskins:1985tn}, Washington, Colorado \cite{Long:2003dx}, Stanford
                    \cite{Chiaverini:2002cb}, Bordag \cite{Bordag:2001qi}, Mostepanenko 1 and 2 \cite{Mostepanenko:2001fx}.}
                \label{fig:fifth_force}
        \end{figure}

        Our interaction potential \eqref{eq:fifth:final_potential} has already been cast in the traditional form for ISL tests
        \begin{equation}
            V(r)=-\frac{GM_1M_2}{r}\Big(1+\alpha \ee^{-r/\lambda}\Big)\ ,
        \end{equation}
        so we can easily adapt the existing constraints to our case $\alpha=\frac{4}{3}\frac{G_h}{G}$, $\lambda=1/m$. The relevant bounds are shown in
        Figure \ref{fig:fifth_force}, for solar-system and laboratory constraints, respectively. We now briefly summarize the content of the experiments
        quoted and refer the reader to the original references and topical reviews \citep{Adelberger:2003zx,Adelberger:2009zz} for further
        details.
        
        \begin{enumerate}
            \item \emph{Planetary} ($10^{9}$ - $10^{13}$ m). One of the effects produced by a modification of the ISL over solar-system scales is an
                anomalous precession of planetary orbits. This fact was used in \cite{Talmadge:1988qz} to set bounds on possible modifications of
                Newtonian gravity, analyzing the orbits of Mercury and Mars.
            \item \emph{Earth-LAGEOS-Moon} ($10^{5}$ - $10^{10}$ m). The first of the curves (LLR) corresponds to a measure of the anomalous precession
                of the Moon, which is the same effect as in the previous point. The other two correspond to measurements of the spatial variation of $G$, based on 
                the orbits of the Moon and the LAGEOS satellite (in an orbit of about $1.2\times 10^{7}$ m).  More details in \cite{DeRujula:1986ug}.
            \item \emph{Geophysical} ($1$ - $10^{4}$ m). There are several experiments halfway between solar-system and laboratory scales, which aim to
                measure spatial variations of $G$ within the Earth. These include measurements in towers, seas, mines and are reviewed in
                \cite{Stacey:1987qj}.
            \item \emph{Cavendish} ($10\ \mu\text{m}$ - $1$ cm). In this range lie the laboratory probes of the force of gravity with torsion balances.
                For a review, see \cite{Adelberger:2003zx}.
            \item \emph{Casimir} ($1$ nm - $10\ \mu$m). Although experimentally challenging, it is possible to measure the Casimir force between two
                bodies, e.g. using atomic-force microscopes. As reviewed in \cite{Mostepanenko:2001fx}, these measurements can be used to 
                constrain the existence of a new force.
        \end{enumerate}
        
        The tightest constraints on the interaction strength come from experiments testing large distances and put, in its turn, strong constraints on the 
        existence of very low mass particles. The situation is reversed for higher masses. In view of the huge experimental challenges, the Casimir 
        experiments, that probe the shortest distances, set significantly looser bounds than its Cavendish counterparts.
        
        The shortest range experiments in the laboratory can only put bounds on masses of about few eVs, and there are no prospects that they can go much
        further. It is in this range of masses where we need the information provided by astrophysical objects.
        
\section{\label{sec:astro}Astrophysical constraints}        
    Stars have become one of the best laboratories to study the impact of new light and weakly interacting particles. One of the main advantages of     
    stars is that, owing to its big size, even really weakly interacting particles can be copiously produced and have a dramatic impact on the
    stellar life. Among the disadvantages, the results are never as statistically significant as in the laboratory experiments, with
    the errors being dominated by astrophysical uncertainties.

    The work in this section can be thought as three different tasks: i) choose the interaction and identify the relevant processes, ii) compute the matrix
    element for each process and the associated energy-loss rate, iii) apply the results to different stellar-medium conditions and compare with 
    observational data. 
    
    \subsection{Interaction}
        The coupling of the hidden graviton is taken to have the same form as the standard graviton, but suppressed by a different energy scale, 
        $\kappa=1/M_h=\sqrt{8\pi G_h}$. It will couple to matter through the energy-momentum tensor obtained with the usual prescription in GR, as the 
        functional derivative with respect to the metric of a minimally coupled matter action. It can be proven \cite{wald2010general} that this 
        prescription gives a suitable symmetric, conserved source.
        
        The most relevant coupling in this work is to QED \cite{Birrell:1982ix}
        \begin{align}
            \mathcal{L}_\text{QED} &= -\frac{1}{4}F_{\mu\nu}F^{\mu\nu}+ \bar{\psi}(\ii\slashed{D}-m)\psi\ ,\qquad D_\mu =\partial_\mu +\ii qA_\mu\ , \\
            T^{\text{QED}}_{\mu\nu} &= F_{\mu\alpha}F^{\n\alpha}_\nu
                    +\frac{\ii}{2}\Big[\bar{\psi}\gamma_{(\mu}D_{\nu )}\psi-(D_{(\mu}\bar{\psi})\gamma_{\nu )}\psi\Big]
                    +\eta_{\mu\nu}\mathcal{L}_\text{QED}\ ,
        \end{align}
        where for hidden gravitons on-shell the last term is irrelevant, see \eqref{massive:eom4}. From this we can read three kind of vertices. The Feynman
        rules for these interactions were calculated in \cite{Giudice:1998ck,Han:1998sg}.
        
    \subsection{Processes}
        It is important to note that these processes take place in a hot plasma where all kinds of new effects appear, as summarized in 
        \cite{Raffelt:1990yz} and references therein. The standard vertices and propagators of quantum field theory (QFT) are modified, new degrees
        of freedom appear (like the longitudinal plasmon) and some collective behaviours are relevant.
        However, as a first approximation, we will neglect most of these plasma effects, pointing out some cases where they can suppress decisively
        some processes.        
        %throughout this work we will neglect these plasma effects in our calculations and we will only point out some cases where they can
        %suppress decisively some processes.       
        %To sum up, we will compute thermally-averaged cross sections using zero temperature QFT plus kinetic theory by 
        %using the Boltzmann equation.
        To sum up, we will use the Boltzmann equation, computing thermally-averaged cross sections with zero temperature QFT.
        
        The Boltzmann equation describes the evolution of the distribution function for different coupled particles  
        \cite{Liddle:2000cg}
        \begin{align}
            \frac{\di f}{\di t} &=\mathcal{C}[f]\ ,\\
            n(t) &=\frac{g}{(2\pi)^{3}}\int f(E,t)\di^{3}\bm{p}\ ,
        \end{align}
        where $n(t)$ is the number density of particles, $g$ is the number of internal degrees of freedom and $\mathcal{C}[f]$ is the collision term. 
        For instance, for processes $ab\leftrightarrow cd$, the collission term for the species $a$ is 
        \begin{align*}
            \mathcal{C}[f_a]=\frac{S}{2E_a}\int &\text{d}\Pi_b\text{d}\Pi_c\text{d}\Pi_d (2\pi)^{4}\delta^{4}(p_a+p_b-p_c-p_d)|\mathcal{M}|^{2}\\
                &\times \Big[\underbrace{f_cf_d(1\pm f_a)(1\pm f_b)}_{\displaystyle cd\to ab}-\underbrace{f_af_b(1\pm f_c)(1\pm 
                    f_d)}_{\displaystyle ab\to cd}\Big]\left\{\begin{array}{l}+\ \text{bosons}\\ -\ \text{fermions}\end{array}\right.
        \end{align*}
        where $\text{d}\Pi_i = \frac{\text{d}\bm{p}^{3}_i}{(2\pi)^{3}2E_i}$ is the Lorentz-invariant phase-space volume and $S$ is the proper symmetry factor, 
        e.g. $S=1/2$ for identical particles in the initial or final state. The energy-loss rate, i.e. energy released per unit volume and per unit of time,
        due to emission of $a$-particles is
        \begin{align}\label{eq:Boltzman_2_2}
            Q_a &=S\int E_a \text{d}\Pi_a \int \text{d}\Pi_b \text{d}\Pi_c \text{d}\Pi_d (2\pi)^{4}\delta^{4}(p_a+p_b-p_c-p_d)
                f_cf_d(1\pm f_b) \sum_\text{spins}|\mathcal{M}|^{2}\ ,
        \end{align}
        assuming that the particles are readily emitted, so we neglect the backreaction $ab\to cd$ and the enhancement/blocking 
        factor $(1\pm f_a)$. Of course, we also consider thermal equilibrium, so the $f$'s are the equilibrium Bose-Einstein/Fermi-Dirac distributions.
        Using similar arguments, if we have a process with only one particle in the final state $cd\to a$, the energy released is
        \begin{equation}\label{eq:Boltzman_2_1}
            Q_a = S\int f_c\text{d}\Pi_c \int f_d\text{d}\Pi_d \frac{E_a}{2\sqrt{s}}(2\pi)\delta(m_a-\sqrt{s})\sum_\text{spins}|\mathcal{M}|^{2}\ ,
        \end{equation}
        where $s=-(p_c+p_d)^{2}$ is associated with the square of the center of mass energy. In general, the Mandelstam variables are defined as
        \begin{align}
            s&=-(p_c+p_d)^2\ ,\\
            t&=-(p_c-p_a)^2\ ,\\
            u&=-(p_d-p_a)^2\ .
        \end{align}             
        Finally, in the presence of an external field, the momentum is conserved without affecting the conservation of energy. For a process of the type $c+\text{EF}\to ab$, we have
        \begin{equation}\label{eq:Boltzman_1EF_2}
            Q_a = n_NS\int E_a\di\Pi_a \int (1\pm f_b)\di\Pi_b\int f_c\di\Pi_c (2\pi)\delta (E_a+E_b-E_c)\sum_\text{spins} |\mathcal{M}|^{2}\ .
        \end{equation}
        Now, we will analyze the different processes relevant for the emission of hidden gravitons in stars.        
        \subsubsection{Photon-photon}
            The amplitude for this process is
            \begin{equation}
                \vcenter{\hbox{\begin{fmffile}{dia_gg}
                    \begin{fmfgraph*}(90,70)
                        \fmfleft{i1,i2}
                        \fmfright{o1}
                        \fmflabel{$h_{\mu\nu}$}{o1}     
                        
                        \fmf{photon,label=$A_\alpha (k_1)$,l.s=left,l.d=0.04w}{i2,v1}
                        \fmf{photon,label=$A_\beta (k_2)$,l.s=right,l.d=0.04w}{i1,v1}
                        \marrow{a}{left}{bot}{$ $}{i1,v1}
                        \marrow{b}{left}{bot}{$ $}{i2,v1}
                        \fmf{dbl_wiggly}{v1,o1}
                        \fmfv{decor.shape=circle,decor.filled=full,decor.size=.05w}{v1}
                    \end{fmfgraph*}
                \end{fmffile}}}\qquad = \quad -2\ii\kappa \epsilon^{\alpha}(k_1)\epsilon^{\beta}(k_2)
                    V_{(\mu\nu)\alpha\beta}(k_1,k_2)\epsilon^{\mu\nu}(k_1+k_2)\ .
            \end{equation}
            where the vertex is
            \begin{equation}
                V_{\mu\nu\alpha\beta}(k_{1}, k_{2}) = -\frac{1}{2}\eta_{\mu \nu}(k_{1\beta}k_{2\alpha}-\eta_{\alpha\beta}k_{1}\cdot k_{2})
                    -\eta_{\alpha\beta}k_{1 \mu}k_{2\nu} -\eta_{\mu\alpha}(\eta_{\nu\beta}k_{1}\cdot k_{2}-k_{1\beta}k_{2\nu})
                    +\eta_{\mu\beta}k_{1\nu}k_{2\alpha}\ .
            \end{equation}
            Summing over initial and final spins, we can easily obtain the matrix element 
            \begin{equation}
                \sum_\text{spins}|\mathcal{M}|^{2} = 4\kappa^{2}S^{\mu\nu\mu'\nu'}V_{\mu\nu\alpha\beta}V_{\mu'\nu'}^{\n\n\n\alpha\beta} = 
                    -8\kappa^{2}(k_1k_2)^{2} = 2\kappa^{2}s^{2}\ .
            \end{equation}
            Now we can plug it in our Boltzmann equation \eqref{eq:Boltzman_2_1}, with the appropiate Bose-Einstein distributions and a symmetry 
            factor $S=1/2$ for identical particles in the initial state, to compute the energy loss
            \begin{align}\label{eq:Q_gg}
                Q_\gamma &=\frac{S}{4(2\pi)^{3}}\int f_cf_d\ p_c\di E_c\ p_d\di E_d\ \di z_{cd}\frac{E_c+E_d}{m}\delta (m-\sqrt{s})\sum_\text{spins}
                        |\mathcal{M}|^{2}\nonumber\\
                    &= -\frac{\kappa^{2}m^{4}T^{3}}{2(2\pi)^{3}}\int^{\infty}_0\frac{\omega\di\omega}{\ee^{\omega}-1}\log\Big(1-
                        \ee^{-\frac{m^{2}}{4T^{2}\omega}}\Big)\ ,
            \end{align}
            where $s=2\omega_c\omega_d(1-z_{cd})$ is the center of mass energy and $z_{cd}\equiv \cos(\theta_{cd})$ is the cosine of the angle between 
            the incident photons.       
        \subsubsection{Gravi-Compton}
            The Gravi-Compton process consists on four diagrams, Figure \ref{fig:process_compton}.
            \begin{figure}[!htb]
                \caption{Gravi-Compton process}
                \label{fig:process_compton}
                \subfigure{
                    \begin{fmffile}{dia_gcp1}
                        \begin{fmfgraph*}(90,70)
                            \fmfleft{i1,i2}
                            \fmfright{o1,o2}
                            \fmf{photon,label=$A_\alpha (k)$,l.s=right,l.d=0.04w}{i2,v1}
                            \fmf{fermion,label=$f(p)$,l.s=left,l.d=0.04w}{i1,v1}
                            \fmf{fermion}{v1,v2}
                            \fmf{dbl_wiggly,label=$h_{\mu\nu}(q)$,l.s=right,l.d=0.04w}{v2,o2}
                            \fmf{fermion,label=$f(p')$,l.s=left,l.d=0.04w}{v2,o1}
                            \fmfv{decor.shape=circle,decor.filled=full,decor.size=.05w}{v1}
                            \fmfv{decor.shape=circle,decor.filled=full,decor.size=.05w}{v2}
                        \end{fmfgraph*}
                    \end{fmffile}
                }
                \subfigure{
                    \begin{fmffile}{dia_gcp2}
                        \begin{fmfgraph*}(90,70)
                            \fmfleft{i1,i2}
                            \fmfright{o1,o2}
                            \fmf{fermion}{i2,v1}
                            \fmf{photon}{i1,v2}
                            \fmf{fermion}{v1,v2}
                            \fmf{dbl_wiggly}{v1,o2}
                            \fmf{fermion}{v2,o1}
                            \fmfv{decor.shape=circle,decor.filled=full,decor.size=.05w}{v1}
                            \fmfv{decor.shape=circle,decor.filled=full,decor.size=.05w}{v2}
                        \end{fmfgraph*}
                    \end{fmffile}
                }
                \subfigure{
                    \begin{fmffile}{dia_gcp3}
                        \begin{fmfgraph*}(90,70)
                            \fmfleft{i1,i2}
                            \fmfright{o1,o2}
                            \fmf{photon}{i2,v1}
                            \fmf{fermion}{i1,v2}
                            \fmf{photon}{v1,v2}
                            \fmf{dbl_wiggly}{v1,o2}
                            \fmf{fermion}{v2,o1}
                            \fmfv{decor.shape=circle,decor.filled=full,decor.size=.05w}{v1}
                            \fmfv{decor.shape=circle,decor.filled=full,decor.size=.05w}{v2}
                        \end{fmfgraph*}
                    \end{fmffile}
                }
                \subfigure{
                    \begin{fmffile}{dia_gcp4}
                        \begin{fmfgraph*}(90,70)
                            \fmfleft{i1,i2}
                            \fmfright{o1,o2}
                            \fmf{photon}{i2,v1}
                            \fmf{fermion}{i1,v1}
                            \fmf{dbl_wiggly}{v1,o2}
                            \fmf{fermion}{v1,o1}
                            \fmfv{decor.shape=circle,decor.filled=full,decor.size=.05w}{v1}
                        \end{fmfgraph*}
                    \end{fmffile}
                }
            \end{figure}
            The scattering amplitude is
            \begin{equation}
                \ii\mathcal{M} \equiv -\ii\kappa e\Big(\mathcal{A}_{(\mu\nu)\alpha}^{\text{(I)}}+\mathcal{A}_{(\mu\nu)\alpha}^{\text{(II)}}
                    +\mathcal{A}_{(\mu\nu)\alpha}^{\text{(III)}}+\mathcal{A}_{(\mu\nu)\alpha}^{\text{(IV)}}\Big)\epsilon^{\mu\nu}(q)\epsilon^{\alpha}(k)\ ,
            \end{equation}
            where
            \begin{align}
                \mathcal{A}_{\mu\nu\alpha}^{\text{(I)}} &= \bar{u}\frac{p_\mu+k_\mu-q_\mu/2}{(p+k)^{2}+m_e^{2}}\gamma_\nu(\slashed{p}+\slashed{k}-m_e)\gamma_\alpha u
                    +\bar{u}\frac{\eta_{\mu\nu}}{(k+p)^{2}+m_e^{2}}(\slashed{k}+\slashed{p}-\slashed{q}/2+2m_e)(\slashed{k}+\slashed{p}-m_e)
                    \gamma_\alpha u\ , \nonumber\\
                \mathcal{A}_{\mu\nu\alpha}^{\text{(II)}} &= \bar{u}\frac{p_\mu-q_\mu/2}{(p-q)^{2}+m_e^{2}}\gamma_\alpha(\slashed{p}-\slashed{q}-m_e)\gamma_\nu u
                    +\bar{u}\frac{\eta_{\mu\nu}}{(p-q)^{2}+m_e^{2}}\gamma_\alpha(\slashed{p}-\slashed{q}-m_e)(\slashed{p}-\slashed{q}/2+2m_e)u\ ,\nonumber\\
                \mathcal{A}_{\mu\nu\alpha}^{\text{(III)}} &= \frac{2}{(q-k)^{2}}\bar{u}\gamma^{\beta}V_{\mu\nu\beta\alpha}(q-k,k)u\ ,\nonumber\\
                \mathcal{A}_{\mu\nu\alpha}^{\text{(IV)}} &= \bar{u}(\gamma_\mu \eta_{\nu\alpha}-\eta_{\mu\nu}\gamma_\alpha)u\ .\nonumber           
            \end{align}
            The squared matrix element, summing over spins, is
            \begin{align}\label{eq:compton_F}
                \sum_\text{spins} |\mathcal{M}|^{2} &= (\kappa e)^{2}S^{\mu\nu\mu'\nu'}\text{Tr}\Big[\mathcal{A}_{\mu\nu\alpha}
                        (\slashed{p}-m_e)\bar{\mathcal{A}}_{\mu'\nu'}^{\n\n\n\alpha}(\slashed{p'}-m_e)\Big]\nonumber\\
                    &=(\kappa e)^{2}F(s,t)\ ,
            \end{align}
            where $F(s,t)$ is a fairly lengthy function of the Mandelstam variables $s$, $t$ and the masses of the particles, that we will integrate
            numerically later on. 
            
            The final result for the process $\gamma(c)+e(d)\to e(b)+G(a)$ is
            \begin{align}\label{eq:Q_compton}
                Q_\text{cp}= \frac{\kappa^{2}e^{2}}{8(2\pi)^{5}}
                        \int^{\infty}_0 &\frac{E_c\di E_c}{\ee^{E_c/T}-1}\int^{\infty}_{m_e}\frac{p_d\di E_d}{\ee^{(E_d-\mu)/T}+1}\int^{1}_{-1}\di z_{cd}
                        \int^{1}_{-1}\di z_\text{cm}\Big(1-f_\text{F}(E_b)\Big)\nonumber\\
                    &\times\frac{p_\text{cm}E_a}{\sqrt{s}}\theta(\sqrt{s}-m-m_e) F(s,t)\ ,
            \end{align}
            where $E_c$ and $E_d$ are the energy of the initial particles, $z_{cd}\equiv \cos(\theta_{cd})$ is the angle between these initial particles,
            $z_\text{cm}\equiv\cos(\theta_\text{cm})$ is the angle between the initial and final particles in the center of mass (CM) frame and
            $p_\text{cm}$ is the momentum of the final particles in this CM frame
            \begin{equation}
                p_\text{cm} = \frac{1}{2\sqrt{s}}\sqrt{s^{2}-2(m^{2}+m_e^{2})s+(m^{2}-m_e^{2})^{2}}\ .
            \end{equation}
            Finally, $E_a$ and $E_b$ are the energies of the final states in an arbitrary frame, that can be obtained from its CM value $E_\text{cm}=
            \sqrt{m^{2}+p_\text{cm}^{2}}$ with a boost.
        \subsubsection{Electron-positron annihilation}
            There are two electron-positron processes that are important, $e^+e^-\to G$ and $e^-e^+\to\gamma G$. We will make an important 
            approximation throughout this subsection. 
            Since the amount of positrons in the red giants and the Sun is negligible, this process will only be important in supernovas. But, in that 
            case, the electrons are highly relativistic, $m_e\ll T_{\text{SN}}$, so we can safely set $m_e\simeq 0$ in our calculations.
            
            The first process is equivalent to the photon-photon annihilation
            \begin{equation}
                \vcenter{\hbox{\begin{fmffile}{dia_ee1}
                    \begin{fmfgraph*}(90,70)
                        \fmfleft{i1,i2}
                        \fmfright{o1}   
                        
                        \fmf{fermion,label=$f (k_1)$,l.s=left,l.d=0.04w}{i2,v1}
                        \fmf{fermion,label=$\bar{f} (k_2)$,l.s=left,l.d=0.04w}{v1,i1}
                        \marrow{a}{left}{bot}{$ $}{i1,v1}
                        \marrow{b}{left}{bot}{$ $}{i2,v1}
                        \fmf{dbl_wiggly}{v1,o1}             
                        \fmfv{decor.shape=circle,decor.filled=full,decor.size=.05w}{v1}
                    \end{fmfgraph*}
                \end{fmffile}}}\qquad = \quad -\frac{\ii\kappa}{2}\bar{v}(k_2) W_{(\mu\nu)}(k_1,-k_2)u(k_1)\epsilon^{\mu\nu}(k_1+k_2)\ .
            \end{equation}
            where the vertex is
            \begin{equation}
                W_{\mu\nu}(k_1,k_2) = (k_1 + k_2)_{\mu}\gamma_{\nu} - \eta_{\mu \nu}(\slashed{k_1} + \slashed{k_2} + 2m_e)\ .
            \end{equation}          
            As in the photon-photon case, the matrix element is easily computed
            \begin{align}
                \sum_\text{spins}|\mathcal{M}|^{2} &= \frac{\kappa^{2}}{4}(k_1-k_2)_\mu(k_1-k_2)_{\mu'}S^{\mu'\nu'\mu\nu}\text{Tr}\Big[
                    \gamma_\nu(\slashed{k_1}-m_e)\gamma_{\nu'}(\slashed{k_2}+m_e)\Big]\nonumber\\
                    &= \frac{\kappa^{2}}{2}\Big(s^{2}+\frac{4}{3}m_e^{2}s-\frac{32}{3}m_e^{4}\Big)\simeq \frac{\kappa^{2}}{2}s^{2}\ ,
            \end{align}
            and the corresponding energy-loss rate is
            \begin{equation}\label{eq:Q_ee1}
                Q_{ee1} = \frac{\kappa^{2}m^{4}T^{3}}{8(2\pi)^{3}}\int^{\infty}_0 \frac{E\di E}{\ee^{E+\mu/T}+1}\log\Big(1+
                    \ee^{-\frac{m^{2}}{4T^{2}E}+\mu/T}\Big) \quad+\quad (\mu\to -\mu)\ .
            \end{equation}
            The second process involves a photon and a hidden graviton in the final state, Figure \ref{fig:process_ee2}, so it can also take place in massless 
            gravity. 
            \begin{figure}[htb]
                \caption{Electron-positron annihilation}
                \label{fig:process_ee2}
                \subfigure{
                    \begin{fmffile}{dia_ee2_1}
                        \begin{fmfgraph*}(90,70)
                            \fmfleft{i1,i2}
                            \fmfright{o1,o2}
                            \fmf{fermion,label=$f(p)$,l.s=right,l.d=0.04w}{i2,v1}
                            \fmf{fermion,label=$\bar{f}(p')$,l.s=right,l.d=0.04w}{v2,i1}
                            \fmf{fermion}{v1,v2}
                            \fmf{photon,label=$A_\alpha(k)$,l.s=right,l.d=0.04w}{v1,o2}
                            \fmf{dbl_wiggly,label=$h_{\mu\nu}(q)$,l.s=left,l.d=0.04w}{v2,o1}
                            \fmfv{decor.shape=circle,decor.filled=full,decor.size=.05w}{v1}
                            \fmfv{decor.shape=circle,decor.filled=full,decor.size=.05w}{v2}
                        \end{fmfgraph*}
                    \end{fmffile}
                }
                \subfigure{
                    \begin{fmffile}{dia_ee2_2}
                        \begin{fmfgraph*}(90,70)
                            \fmfleft{i1,i2}
                            \fmfright{o1,o2}
                            \fmf{fermion}{i2,v1}
                            \fmf{fermion}{v2,i1}
                            \fmf{fermion}{v1,v2}
                            \fmf{dbl_wiggly}{v1,o2}
                            \fmf{photon}{v2,o1}
                            \fmfv{decor.shape=circle,decor.filled=full,decor.size=.05w}{v1}
                            \fmfv{decor.shape=circle,decor.filled=full,decor.size=.05w}{v2}
                        \end{fmfgraph*}
                    \end{fmffile}
                }
                \subfigure{
                    \begin{fmffile}{dia_ee2_3}
                        \begin{fmfgraph*}(90,70)
                            \fmfleft{i1,i2}
                            \fmfright{o1,o2}
                            \fmf{fermion}{v1,i1}
                            \fmf{fermion}{i2,v1}
                            \fmf{photon}{v1,v2}
                            \fmf{dbl_wiggly}{v2,o2}
                            \fmf{photon}{v2,o1}
                            \fmfv{decor.shape=circle,decor.filled=full,decor.size=.05w}{v1}
                            \fmfv{decor.shape=circle,decor.filled=full,decor.size=.05w}{v2}
                        \end{fmfgraph*}
                    \end{fmffile}
                }
                \subfigure{
                    \begin{fmffile}{dia_ee2_4}
                        \begin{fmfgraph*}(90,70)
                            \fmfleft{i1,i2}
                            \fmfright{o1,o2}
                            \fmf{fermion}{v1,i1}
                            \fmf{fermion}{i2,v1}
                            \fmf{dbl_wiggly}{v1,o2}
                            \fmf{photon}{v1,o1}
                            \fmfv{decor.shape=circle,decor.filled=full,decor.size=.05w}{v1}
                        \end{fmfgraph*}
                    \end{fmffile}
                }
            \end{figure}
            
            The amplitude and cross section for this case can be adapted from the Compton process \eqref{eq:compton_F} using the crossing symmetry
            \begin{equation}
                \sum_\text{spins}|\mathcal{M}|^{2} = (\kappa e)^{2}F(t,s)\ .
            \end{equation}
            In the limit $m_e\to 0$, the function $F(t,s)$ takes a simple form
            \begin{equation}
                F(t,s) \simeq \frac{\Big(M^{4}-2M^{2}t+s^{2}+2t(s+t)\Big)\Big(4t(s+t)-M^{2}(s+4t)\Big)}{st(s+t-M^{2})}\ .
            \end{equation}
            The final result for the process $\bar{e}(c)+e(d)\to \gamma(b)+G(a)$ is
            \begin{align}\label{eq:Q_ee2}
                Q_{ee2}= \frac{\kappa^{2}e^{2}}{8(2\pi)^{5}}
                        \int^{\infty}_0 &\frac{E_c\di E_c}{\ee^{(E_c+\mu)/T}+1}\int^{\infty}_{0}\frac{E_d\di E_d}{\ee^{(E_d-\mu)/T}+1}\int^{1}_{-1}\di z_{cd}
                        \int^{1}_{-1}\di z_\text{cm}\Big(1+f_\text{B}(E_b)\Big)\nonumber\\
                    &\times\frac{p_\text{cm}E_a}{\sqrt{s}}\theta(\sqrt{s}-m) F(t,s)\ .
            \end{align}
        \subsubsection{Gravi-bremsstrahlung}
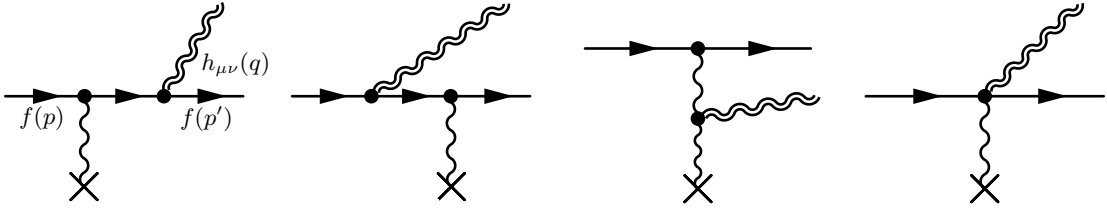
\begin{figure}[!htb]
                \caption{Gravi-bremsstrahlung process}
                \label{fig:process_bremss}
                \subfigure{
                    \begin{fmffile}{dia_gb1}
                        \begin{fmfgraph*}(90,70)
                            \fmfleft{i0,i1,i2}
                            \fmfbottom{b1,b2,b3,b4}
                            \fmfright{o0,o1,o2}
                            \fmf{fermion,label=$f(p)$,l.s=right,l.d=0.04w}{i1,v1}
                            \fmf{fermion}{v1,v2}
                            \fmf{fermion,label=$f(p')$,l.s=right,l.d=0.04w}{v2,o1}
                            \fmffreeze
                            \fmf{photon,tension=0}{b2,v1}                                                                                   
                            \fmf{dbl_wiggly,label=$h_{\mu\nu}(q)$,l.s=right,l.d=0.04w}{v2,o2}                           
                            \fmfv{decor.shape=circle,decor.filled=full,decor.size=.05w}{v1}
                            \fmfv{decor.shape=circle,decor.filled=full,decor.size=.05w}{v2}
                            \fmfv{decor.shape=cross}{b2}
                        \end{fmfgraph*}
                    \end{fmffile}
                }
                \subfigure{
                    \begin{fmffile}{dia_gb2}                        
                        \begin{fmfgraph*}(90,70)
                            \fmfleft{i0,i1,i2}
                            \fmfbottom{b1,b2,b3,b4}
                            \fmfright{o0,o1,o2}
                            \fmf{fermion}{i1,v1}
                            \fmf{fermion}{v1,v2}
                            \fmf{fermion}{v2,o1}
                            \fmffreeze
                            \fmf{photon,tension=0}{b3,v2}   
                            \fmf{dbl_wiggly}{v1,o2}                 
                            \fmfv{decor.shape=circle,decor.filled=full,decor.size=.05w}{v1}
                            \fmfv{decor.shape=circle,decor.filled=full,decor.size=.05w}{v2}
                            \fmfv{decor.shape=cross}{b3}
                        \end{fmfgraph*}
                    \end{fmffile}
                }
                \subfigure{
                    \begin{fmffile}{dia_gb3}
                        \begin{fmfgraph*}(90,70)
                            \fmfleft{i0,i1,i2,i3,i4}
                            \fmfbottom{b1,b2,b3}
                            \fmfright{o0,o1,o2,o3,o4}
                            \fmf{fermion}{i3,v1}                            
                            \fmf{fermion}{v1,o3}
                            \fmffreeze
                            \fmf{photon,length=0.5}{v1,v2}
                            \fmf{photon,length=0.5}{b2,v2}
                            \fmffreeze                          
                            \fmf{dbl_wiggly}{v2,o2}                                                     
                            \fmfv{decor.shape=circle,decor.filled=full,decor.size=.05w}{v1}
                            \fmfv{decor.shape=circle,decor.filled=full,decor.size=.05w}{v2}
                            \fmfv{decor.shape=cross}{b2}
                        \end{fmfgraph*}
                    \end{fmffile}
                }
                \subfigure{
                    \begin{fmffile}{dia_gb4}
                        \begin{fmfgraph*}(90,70)
                            \fmfleft{i0,i1,i2}
                            \fmfbottom{b1,b2,b3}
                            \fmfright{o0,o1,o2}
                            \fmf{fermion}{i1,v1}
                            \fmf{fermion}{v1,o1}
                            \fmffreeze
                            \fmf{photon,tension=0}{b2,v1}                                                                                   
                            \fmf{dbl_wiggly}{v1,o2}                         
                            \fmfv{decor.shape=circle,decor.filled=full,decor.size=.05w}{v1}
                            \fmfv{decor.shape=cross}{b2}
                        \end{fmfgraph*}
                    \end{fmffile}
                }
            \end{figure}

            For this process we can adapt the result \eqref{eq:compton_F}. Now the photon is off-shell, see Figure \ref{fig:process_bremss}, it is a Coulomb 
            field produced by a static heavy nucleus.
            
            In the external field approximation, we must substitute the polarization vector $\epsilon^{\alpha}(k)$ 
            with the external Coulomb field
            $A^{\alpha}=-\eta^{0\alpha}Ze/k^{2}$, $k^{\mu}=(0,\ \bm{q}+\bm{p}'-\bm{p})$. We are also neglecting the emission of hidden gravitons
            from the nucleus, since its contribution is strongly suppressed by its large mass. The matrix element is
            \begin{align}
                \sum_\text{spins} |\mathcal{M}|^{2} &= \frac{(\kappa Ze^{2})^{2}}{k^{4}} S^{\mu\nu\mu'\nu'}\text{Tr}\Big[\mathcal{A}_{\mu\nu 0}
                    (\slashed{p}+m_e)\bar{\mathcal{A}}_{\mu'\nu' 0}(\slashed{p'}+m_e)\Big]\nonumber\\
                &= (\kappa Ze^{2})^{2}M(s,t,u)\ ,
            \end{align}
            where $M(s,t,u)$ is a lengthy, rational function of the Mandelstam variables and the masses of the particles. In the end, the energy-loss rate
            for the process $e(c)+\text{EF}\to e(b)+G(a)$ is
            \begin{align}\label{eq:Q_gb}
                Q_\text{gb}=\frac{\kappa^{2}e^{4}}{4(2\pi)^{5}}\sum_j Z_j^{2}n_j\int^{\infty}_{m_e}
                    &\di E_b\int^{\infty}_{m_e}\di E_c\theta (E_c-E_b-m)\int^{1}_{-1}\di z_a\int^{1}_{-1}\di z_c p_bp_c(E_c-E_b)\nonumber\\
                    &\times\sqrt{(E_c-E_b)^{2}-m^{2}}f_\text{F}(E_c)(1-f_\text{F}(E_b))M(s,t,u)\ ,
            \end{align}
            where we have summed over all the different nuclei present in the medium. If we assume that the star only contains fully ionized 
            hydrogen and helium,
            \begin{equation}
                \sum_j Z_j^{2}n_j = \sum_j Z_j^{2}\frac{X_j\rho}{A_j m_u} = \frac{\rho}{m_u}\ ,
            \end{equation}
            where $Z_j$ is the atomic number of the element $j$, $X_j$ is the mass fraction, $A_j$ is the atomic weight and $m_u$ is the atomic mass unit.           
            This should be a fair approximation, but it may underestimate the energy production in stars with appreciable metallicity. The heaviest
            nuclei, even in small amounts, can contribute significantly to this mechanism, for they also have higher charge $Z$.
            
        \subsubsection{Nucleon bremsstrahlung}
            As mentioned in the introduction, most of the previous work on astrophysical constraints with massive gravitons was motivated by the ADD
            proposal \citep{ArkaniHamed:1998nn}. Shortly after, these authors studied the phenomenological consequences of the model in     
            \citep{ArkaniHamed:1998rs} and, using order-of-magnitude estimates, pointed out the relevance of two-nucleon processes $N+N\to N+N+G$, 
            in supernovae.
            
            Since then, considerable efforts have been devoted to detailed calculations of this energy-loss mechanism. In \cite{Cullen:1999hc} and
            \cite{Barger:1999jf} the authors adopted a derivative and a Yukawa coupling for the nucleon-pion interaction, respectively, and computed
            the energy-loss rate relying on the one-pion-exchange approximation for the nucleon-nucleon scattering.
            
            An alternative approach was adopted in \cite{Hanhart:2000er}, where the authors dropped the one-pion-exchange approximation and used some
            low-energy theorems to set bounds in a model-independent way. The main assumptions in this case were that the emitted gravitons are soft and
            that the emission rate is dominated by two-body collisions. In this soft limit, the energy of the hidden graviton is much smaller than the other
            scales and it is possible to separate the details of the nucleon-nucleon scattering from the emission process. This result allows to use the
            measured nucleon-nucleon scattering cross-section and dramatically simplifies the calculations.
            
            Finally, it is worth mentioning the results obtained in \cite{Hannestad:2003yd}, where the authors derived some semiclassical formulas for 
            the emission and absorption of hidden gravitons in a nuclear medium, such as a supernova or a neutron star.
                        
            For this work, we will quote the results of \cite{Hanhart:2000er} for a single hidden graviton in a neutron gas (neutron-proton and
            proton-proton processes are subdominant). The energy emitted, in a nuclear bremsstrahlung process in the form of soft hidden gravitons, is
            \begin{align}\label{eq:Q_nb}
                Q_\text{nb} = & S\frac{2^{15/2}G_h M^{9/2}T^{13/2}}{5\pi^{6}}\int^{\infty}_\delta\di u_r\int^{1}_{-1}\di(\cos(\theta)) \int^{\infty}_0
                        \di u_P \int^{u_r-\delta}_{0}\di u'_r\int^{1}_{-1}\di(\cos \theta')\nonumber \\
                    & u_r^{1/2}u_P^{1/2}{u'}^{1/2}_r\bar{u}^{2}\xi[\delta/(u_r-u'_r)]f_1f_2(1-f'_1)(1-f_2')\int^{2\pi}_0\frac{\di \phi}{2\pi}\sin^{2}
                        \theta_\text{cm} |\mathcal{A}(\theta_\text{cm},2T\bar{u})|^{2}\ ,
            \end{align}
            where $S=1/4$ is the symmetry factor in this case, $M$ is the neutron mass, $T$ is the temperature of the neutron gas, $\mu=yT$ is the chemical 
            potential and $m=2T\delta$ is the hidden graviton mass. Other definitions are
            \begin{align}
                f_i &=\frac{1}{\ee^{(u_i-y_i)}+1}\ ,\qquad u_{1,2}=u_P +u_r\pm 2\sqrt{u_Pu_r}\cos\theta\ ,\\
                f'_i &=\frac{1}{\ee^{(u'_i-y_i)}+1}\ ,\qquad u'_{1,2}=u_P +u'_r\pm 2\sqrt{u_Pu'_r}\cos\theta'\ ,\\
                \bar{u} &=(u_r+u_r')/2\ ,\\
                \xi [x] &=\sqrt{1-x^{2}}\Big(\frac{19}{18}+\frac{11}{9}x^{2}+\frac{2}{9}x^{4}\Big)\ ,\\
                \cos\theta_\text{cm} &=\cos\theta\cos\theta' +\sin\theta \sin\theta'\cos\phi\ .
            \end{align}
            Moreover, in the region of interest there is a weak dependence of the neutron-neutron scattering cross-section on the angle and the energy,
            so we can use the approximate result
            \begin{equation}
                \frac{M^{2}|\mathcal{A}|^{2}}{32\pi}\simeq \sigma_0 =25\ \text{mb}\ .
            \end{equation}
            
            The formula \eqref{eq:Q_nb} is strictly valid only when the emitted hidden gravitons are soft ($\omega \ll \frac{\bar{p}^{2}}{M}\ 
            \to \ \frac{|u_r-u_r'|}{u_r+u_r'}\ll 1$). In particular, it is not valid in our whole range of masses, it works up to 
            $m\sim 100\ \text{MeV}$, but for these high masses the phase-space effects dominate the energy loss, so the results should not be 
            significantly modified.
        
    \subsection{Energy loss argument}
        If there exists a new type of particle, light enough to be thermally produced in stellar objects, depending on its coupling strength it 
        can have two effects:
        \begin{itemize}
            \item \emph{Energy loss}. If the particle interacts weakly enough, so that once produced it can freely escape, it acts like an energy
                sink and modifies the stellar evolution.
            \item \emph{Energy transfer}. If the particle gets trapped and interacts with the medium, it contributes to the energy transfer, modifying
                the stellar structure.
        \end{itemize}
        We will focus here in the energy loss argument. In general terms, the presence of a new energy sink makes the star burn the nuclear fuel at
        a higher rate, shortening some phases of the stellar evolution. The specific examples to be treated here are:
        \begin{itemize}
            \item \emph{Sun}. In the presence of a new source of energy loss, the Sun would burn its nuclear fuel faster and shine brighter
                \cite{Frieman:1987ui}. This modified luminosity $L_x$ is not directly observable since the solar models are actually fitted to achieve 
                the observed luminosity $L_\odot$, e.g. modifying the amount of helium. However, we can obtain bounds either by imposing that the solar 
                age is not modified too much or that the initial helium fraction has at least the primordial value. Both criteria agree to give a 
                bound \cite{raffelt1996stars}
                \begin{equation}\label{eq:Sun_restriction}
                    L_x < L_\odot\quad \to \quad \epsilon_x <\epsilon_\odot\ ,
                \end{equation}
                where $\epsilon_\odot =1\ \text{erg}\ \text{g}^{-1}\ \text{s}^{-1}$ is the standard emissivity in the Sun and $\epsilon_x$ is the
                emissivity due to the new type of particle. The theoretical luminosities must be evaluated under the conditions of the solar core
                \begin{align*}
                    \rho &= 156\ \text{g}\ \text{cm}^{-3}\ , & n_e &=6.3\times 10^{25}\ \text{cm}^{-3}\ ,\\
                    T &= 1.3\ \text{keV}\ , & X&=0.35\ ,
                \end{align*}
                where $\rho$ is the density in the solar core, $n_e$ the number density of electrons, $T$ the temperature and $X$ the mass fraction
                of hydrogen. The numerical data in this section come either from \cite{raffelt1996stars} or \cite{Raffelt:1990yz}.
            \item \emph{Red Giant Branch}. After depleting the hydrogen in the inner regions, the low mass stars ($M <2M_\odot$) develop a degenerate, 
                inert, helium core and ascend along the red giant branch. The red giant branch ends when the helium ignites and the stars move to the 
                horizontal branch. With additional energy losses the ignition of helium is delayed (or completely prevented in an extreme case). In 
                the light of observations, a simple analytical bound for new energy losses is \cite{raffelt1996stars}
                \begin{equation}\label{eq:RG_restriction}
                    \epsilon_x < 10\ \text{erg}\ \text{g}^{-1}\ \text{s}^{-1}\ ,
                \end{equation}
                to be evaluated at average conditions for the core of a red giant near the helium flash
                \begin{align*}
                    \rho &= 2\times 10^{5}\ \text{g}\ \text{cm}^{-3}\ ,  & n_e &=6\times 10^{28}\ \text{cm}^{-3}\ ,\\
                    T &= 8.6\ \text{keV}\ , & Y_e&=0.5\ ,
                \end{align*}
                where $Y_e$ is the inverse of the ``mean molecular weight'' for the electrons, such that $n_e =Y_e\rho/m_u$.

	        \begin{figure}[htb]
	                \subfigure{
	                    \includegraphics[scale=0.55]{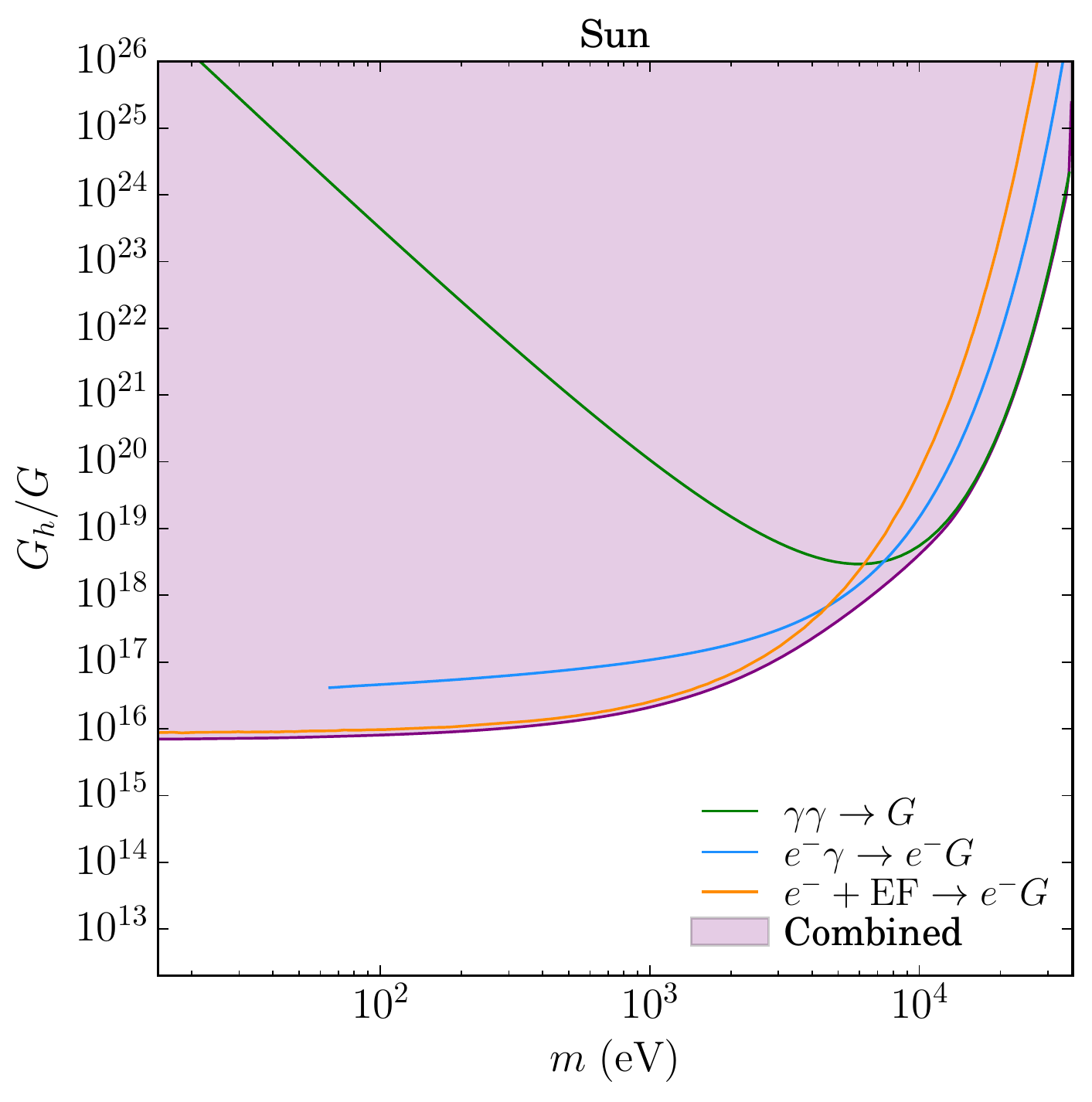}
	                }
	                \subfigure{
	                    \includegraphics[scale=0.55]{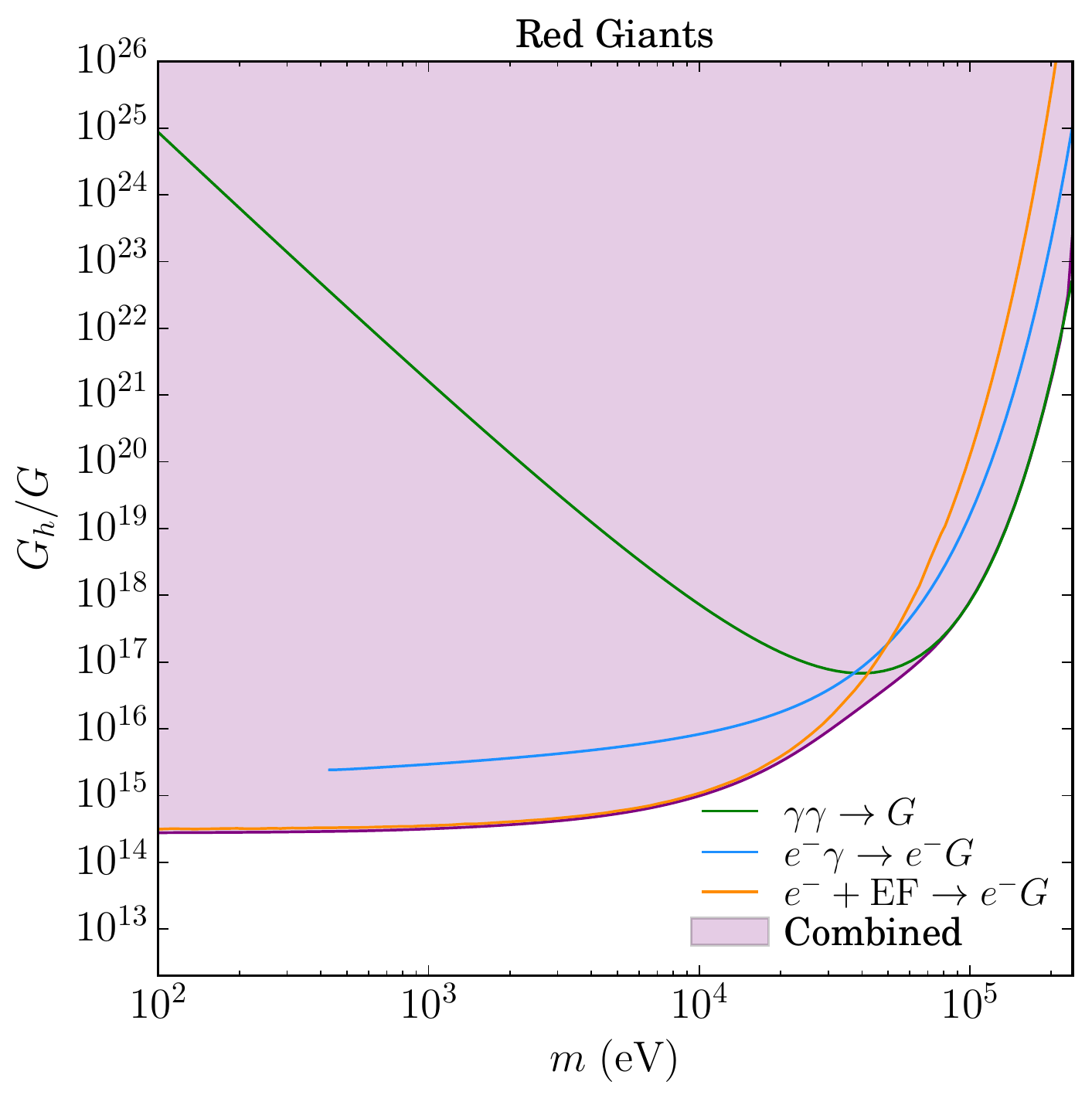}
	                }
	                \subfigure{
	                    \includegraphics[scale=0.55]{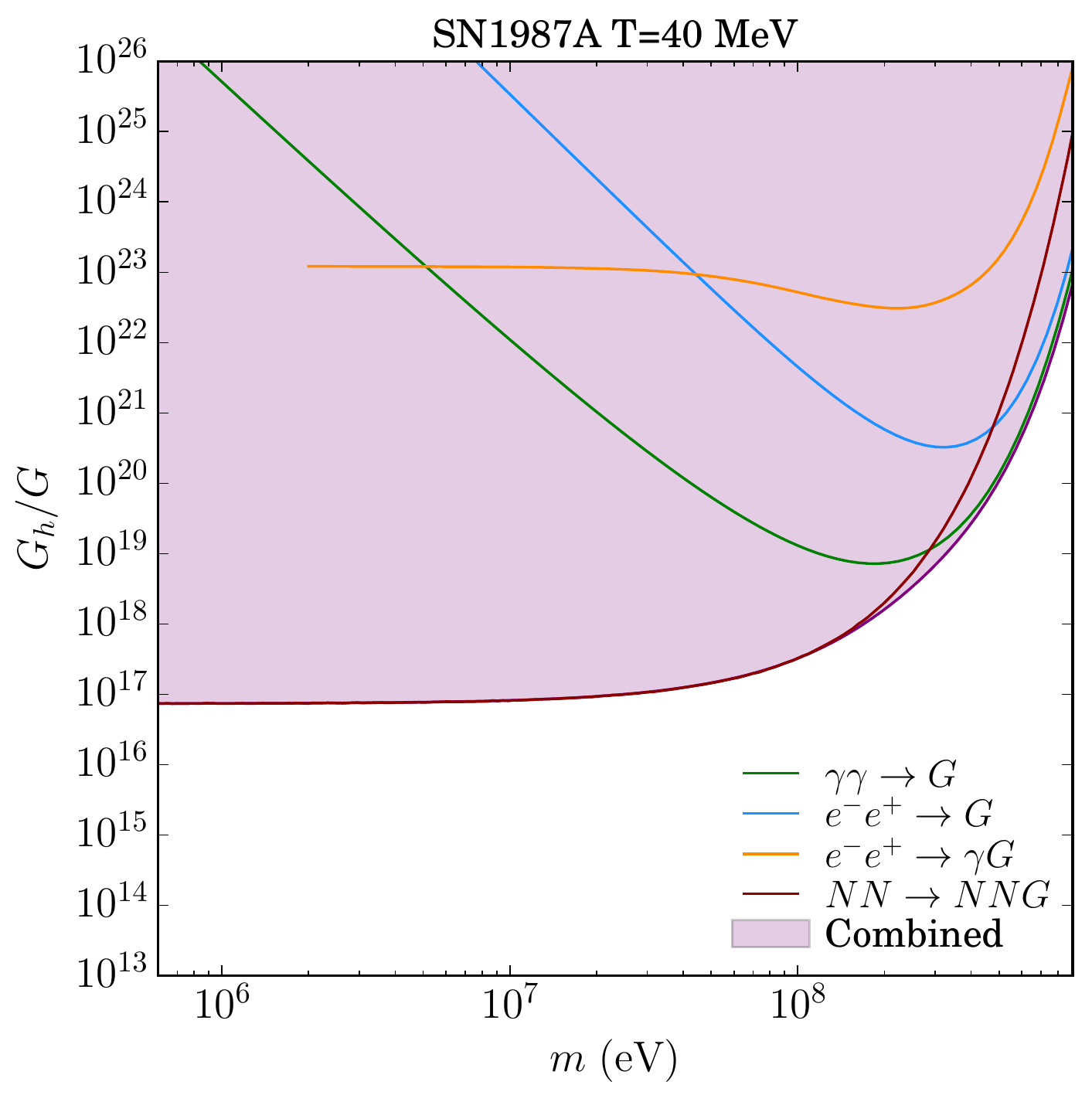}
	                }
	                \subfigure{
	                    \includegraphics[scale=0.55]{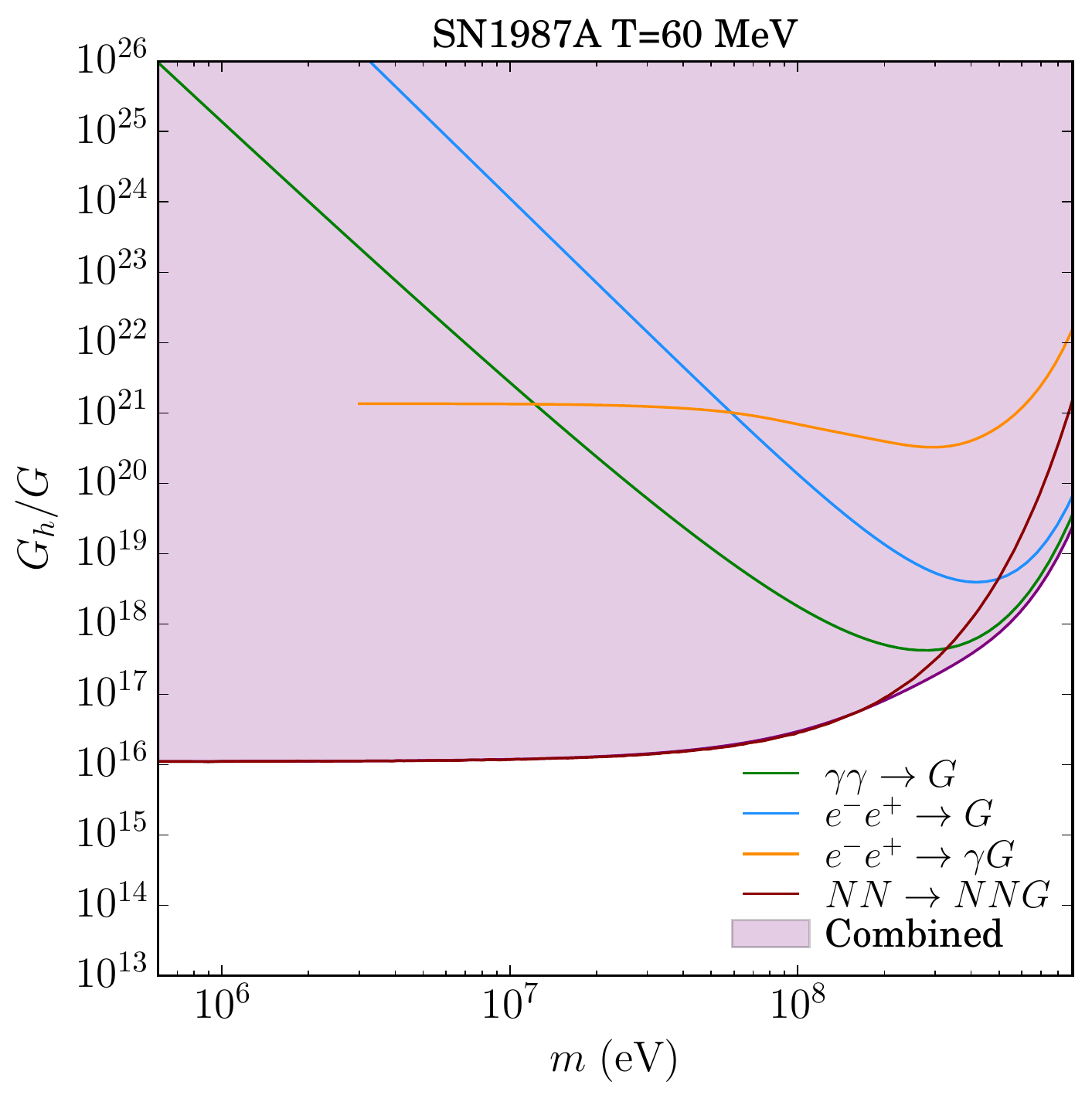}
	                }
	                \caption{Constraints coming from the processes considered under different astrophysical conditions, along with the combined bounds for
	                    each object. The shadowed region is excluded. In the supernova case, as there is uncertainty about its temperature, we plot the 
	                    results for two different temperatures. For the final limits we will use the more conservative estimate of $T=40$ MeV.}
	                \label{fig:astro_tests }
	        \end{figure}
        
            \item \emph{Supernova 1987A}. The energy loss argument for the supernova (SN) case is a bit different from that of standard stars. When a 
                neutron star is born, after a supernova collapse, it emits a huge amount of energy in the form of neutrinos. This is the main cooling 
                mechanism in these objects and any novel form of energy loss would reduce the amount of energy in the form of neutrinos.
                
                The SN1987A event is particularly significant, since the neutrino signal was detected in different observatories around the world. 
                The signal is consistent with the theoretical models, so it can be used to put constraints on the properties of new particles
                that would induce additional energy losses. Raffelt \cite{raffelt1996stars}, based on numerical simulations of SN evolution,  proposed 
                the following analytical criterium
                \begin{equation}\label{eq:SN_restriction}
                    \epsilon_x \lesssim 10^{19}\ \text{erg}\ \text{g}^{-1}\ \text{s}^{-1}\ ,
                \end{equation}
                where it is assumed that the particles escape freely and the energy-loss rate is to be evaluated under conditions
                \begin{align*}
                    \rho &= 8\times 10^{14}\ \text{g}\ \text{cm}^{-3}\ , & T &\sim (40-60)\ \text{MeV}\ .
                \end{align*}
        \end{itemize}

        Finally, all we need to do is to compute the emissivity $\epsilon =Q/\rho$ for each process (\ref{eq:Q_gg}, \ref{eq:Q_compton}, \ref{eq:Q_ee1},
        \ref{eq:Q_ee2}, \ref{eq:Q_gb}, \ref{eq:Q_nb}) under different medium conditions, 
        and apply the restrictions (\ref{eq:Sun_restriction}, \ref{eq:RG_restriction}, \ref{eq:SN_restriction}). The main results are collected in 
        Figure \ref{fig:astro_tests }.

        \begin{figure}[htb]
        \includegraphics[scale=0.9]{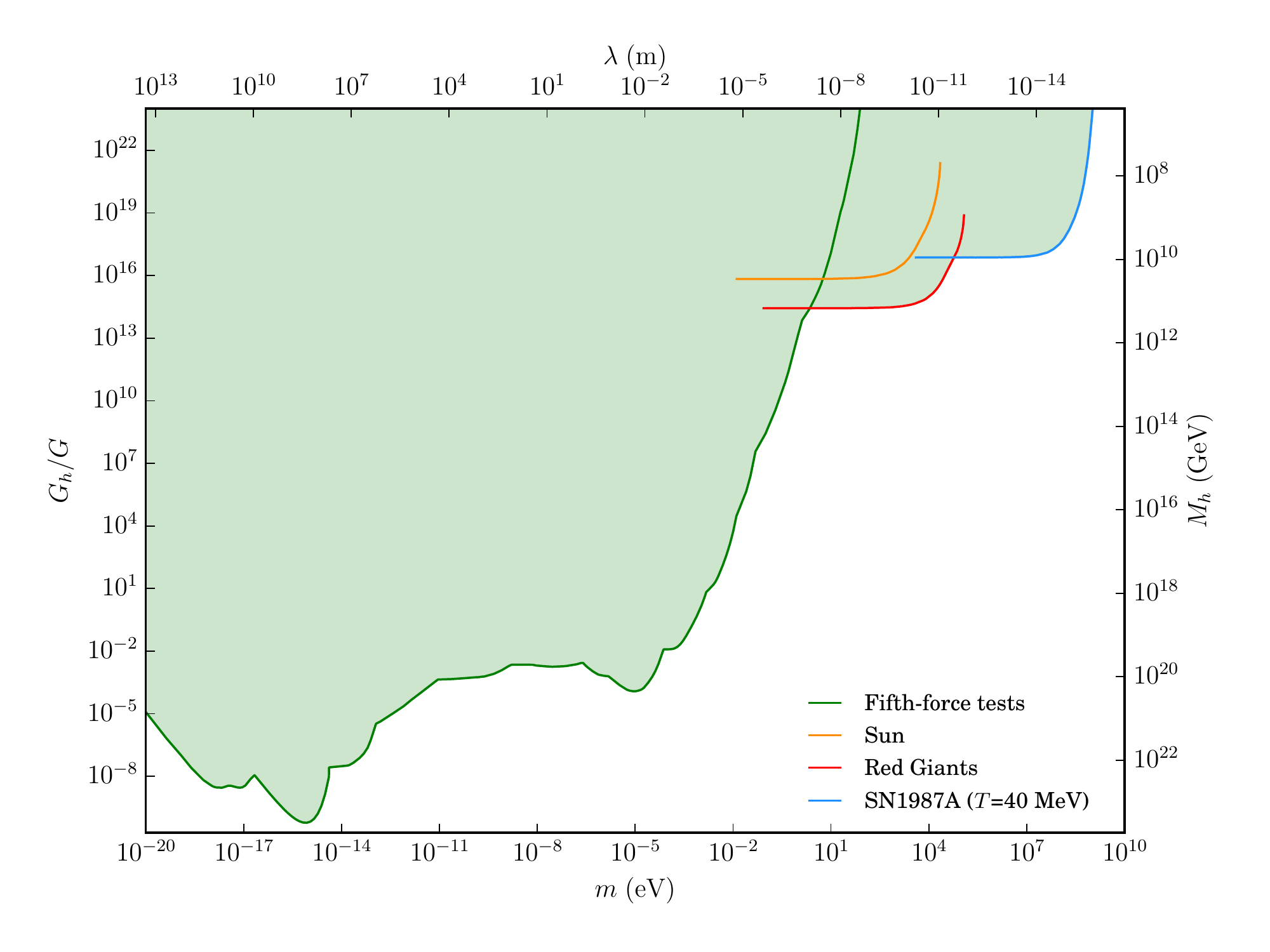}
        \caption{Constraints on the hidden-graviton mass and coupling $G_h$, relative to the standard-graviton coupling. The shadowed 
                    region is excluded by fifth-force tests and energy-loss restrictions, derived in this work. The two additional axes represent the
                    distance scale $\lambda = 1/m$ and the energy scale $M_h = 1/\sqrt{8\pi G_h}$.}
        \end{figure}

    \section{\label{sec:conclusions}Conclusions}

	    In this work, we have derived constraints on the mass and coupling strength of an additional massive graviton. 
	    These new spin-2 particles are a generic feature of different extensions of the gravitational sector. In our analysis, 
	    we have introduced hidden gravitons in the simplest way, as an additional field described by a linear Fierz-Pauli
	    Lagrangian. 
	    In addition to the standard fifth-force
	    tests, we have worked out in detail the emission of these hidden gravitons from different astrophysical objects. The computed emission rates allow us 
	    to place limits on the parameters of the theory, to avoid anomalies in the observed energy-loss rates.    
	    The most important processes in the Sun and red giants are the Compton and the bremsstrahlung process. In the supernova case, these processes are
	    suppressed, since the Pauli blocking is
	    very important and in addition the electric field created by the nucleus is screened, an effect that we have neglected in our calculations. 
	    In this case, there is an appreciable number of positrons in the medium, but their overall contribution to the energy loss turns out to be negligible. 
	    At these nuclear densities, the dominant process is the nucleon-nucleon bremsstrahlung, mediated by the strong interaction.
	    In all three cases the photon-photon process, which is forbidden for massless gravitons, is found to be relevant.
	    
	    These astrophysical bounds complement the fifth-force constraints and are orders of magnitude more competitive than other restrictions in the same range of
	    masses, like tests on atomic systems \cite{Murata:2014nra}. 
	    Further work in this direction would involve a full numerical analysis and a modification of the stellar models. This kind of study has already been 
	    carried out in the case of axions and it would help to refine the constraints and clarify the impact on the stellar structure, as a novel form of energy 
	    transfer for large coupling strengths. 
     
    \vspace{0.2cm}
{\bf Acknowledgements:}
 This work has been supported by the MINECO (Spain) projects FIS2014-52837-P, FIS2016-78859-P(AEI/FEDER, UE),
 and Consolider-Ingenio MULTIDARK CSD2009-00064.

%\clearpage
\bibliography{MiBiblio}

\end{document}